\documentclass[conference]{IEEEtran}
\usepackage{amssymb}
\usepackage[cmex10]{amsmath}
\usepackage{stfloats}
\usepackage{graphicx}
\usepackage{subfigure}
\usepackage{tabularx}
\usepackage{epsfig,epsf,color,balance,cite}
\usepackage{verbatim}
\usepackage{url}
\usepackage{bm}

\usepackage{mathrsfs}
\usepackage{amsthm}
\usepackage{mathtools}
\usepackage{extarrows}
\DeclarePairedDelimiter{\ceil}{\lceil}{\rceil}
\usepackage[a4paper,left=1.3cm,right=1.3cm, top=1.9cm,bottom=2.6cm,textwidth=13.8cm,textheight=20.3cm,
includehead,headheight=0pt,headsep=0pt]{geometry}

\usepackage{algorithm}
\usepackage{algpseudocode}
\usepackage{graphicx}
\usepackage{epstopdf}
\begin{document}

\title{Distributed Information Bottleneck for a Primitive Gaussian Diamond Channel with Rayleigh Fading}

\author{
	\IEEEauthorblockN{Hao Xu\IEEEauthorrefmark{1}\IEEEauthorrefmark{2},
		Kai-Kit Wong\IEEEauthorrefmark{1},
		Giuseppe Caire\IEEEauthorrefmark{2},
		and
		Shlomo Shamai (Shitz)\IEEEauthorrefmark{3}
	}
\IEEEauthorblockA{\IEEEauthorrefmark{1}Department of Electronic and Electrical Engineering, University College London, London WC1E7JE, U.K.}
\IEEEauthorblockA{\IEEEauthorrefmark{2}Faculty of Electrical Engineering and Computer Science, Technical University of Berlin, 10587 Berlin, Germany}
\IEEEauthorblockA{\IEEEauthorrefmark{3}Department of Electrical and Computer Engineering, Technion-Israel Institute of Technology, Haifa 3200003, Israel}
\IEEEauthorblockA{E-mail: hao.xu@ucl.ac.uk; kai-kit.wong@ucl.ac.uk; caire@tu-berlin.de; sshlomo@ee.technion.ac.il}
}

\maketitle

\begin{abstract}
This paper considers the distributed information bottleneck (D-IB) problem for a primitive Gaussian diamond channel with two relays and Rayleigh fading.
Due to the bottleneck constraint, it is impossible for the relays to inform the destination node of the perfect channel state information (CSI) in each realization.
To evaluate the bottleneck rate, we provide an upper bound by assuming that the destination node knows the CSI and the relays can cooperate with each other, and also three achievable schemes with simple symbol-by-symbol relay processing and compression.
Numerical results show that the lower bounds obtained by the proposed achievable schemes can come close to the upper bound on a wide range of relevant system parameters.
\end{abstract}


\IEEEpeerreviewmaketitle

\section{Introduction}
\label{introduction}

Introduced by Tishby in \cite{tishby2000information}, the information bottleneck (IB) paradigm, where relevant  information about 
a signal $X$ is extracted from an observation $Y$ and conveyed to a destination via a rate-constrained bottleneck link, 
has found remarkable applications in communication systems and neural networks \cite{tishby2015deep, shwartz2017opening, goldfeld2020information, zaidi2020information}.
An interesting application of the IB problem in communications consists of a source node, one or more relays, and a destination node, which is connected to the relays via error-free bottleneck links of given rate \cite{winkelbauer2014rate, winkelbauer2014ratevec, 4544988, sanderovich2009distributed, aguerri2019TIT, katz2019gaussian, katz2021filtered, courtade2013multiterminal, estella2018distributed, aguerri2019distributed, caire2018information, info12040155, IBxu}.
Two variants of this problem have been extensively studied. In the information transmission setting, the source wishes to transmit an information message
to the destination. The source-transmitted signal is a codeword, but the relays are ``oblivious'', i.e., unaware of the codebook but only of the marginal statistics 
of the codeword symbols (see \cite{ 4544988} for a rigorous model based on codebook random selection). 
In the (remote) source coding setting, the source produces a random signal with given statistics, and the destination wishes to reproduce it within a certain 
distortion (the so-called Chief Executive Officer (CEO) problem). Interestingly, it turns out that when the distortion is log-loss, the resulting CEO problem has the same achievable 
tradeoff region (relevant information versus bottleneck rates) of the information transmission problem with oblivious relays
(see \cite{4544988} and \cite{estella2018distributed}) although with different operational meaning. 
This tradeoff region is shown to be optimal in \cite{courtade2013multiterminal} for the log-loss CEO problem, and it is shown to 
yield the capacity of the information transmission problem under an additional condition independent of the relay observations condition in \cite{aguerri2019TIT}. 

In both cases,  the source node sends signal sequences over a communication channel and the relays compress and convey their observations to the destination subject to the bottleneck constraints. The ``relevant information'' is expressed by the mutual information between 
the source signal and the messages conveyed by the relays to the destination, and the goal is to maximize such mutual information subject to the bottleneck constraints. 


A brief review of the works in \cite{winkelbauer2014rate, winkelbauer2014ratevec, 4544988, sanderovich2009distributed, aguerri2019TIT, katz2019gaussian, katz2021filtered, courtade2013multiterminal, estella2018distributed, aguerri2019distributed, caire2018information, info12040155, IBxu} is provided 
here in order to put our paper in context. 
References \cite{winkelbauer2014rate} and \cite{winkelbauer2014ratevec} respectively considered Gaussian scalar and vector channels with one relay, and provided the optimal trade-off between the bottleneck and compression rate.
In \cite{4544988, sanderovich2009distributed, aguerri2019TIT, katz2019gaussian, katz2021filtered, courtade2013multiterminal, estella2018distributed, aguerri2019distributed}, 
Tishby's centralized IB method was generalized to the setting with multiple distributed relays and the achievable regions or upper bounds on the capacity were analyzed.
But all references \cite{winkelbauer2014rate, winkelbauer2014ratevec, 4544988, sanderovich2009distributed, aguerri2019TIT, katz2019gaussian, katz2021filtered, courtade2013multiterminal, estella2018distributed, aguerri2019distributed} assumed that the perfect channel state information (CSI) was known at both the relays and the destination node, which is reasonable for block-fading channels, but will be impractical, due to the bottleneck constraint, when channels vary quickly. 
Reference \cite{caire2018information} investigated the IB problem of a scalar Rayleigh fading channel with one relay, where the CSI is only known at the relay.
An upper bound and two achievable schemes which yielded lower bounds to the bottleneck rate were provided by \cite{caire2018information}.
The work was then extended to the vector case by \cite{info12040155} and \cite{IBxu}.

In this paper, we extend the work of \cite{caire2018information} to the primitive Gaussian diamond channel with two relays, each experiencing i.i.d. Rayleigh fading. To evaluate the achievable bottleneck rate (or relevant information), we first obtain an upper bound by assuming that the destination node knows the CSI and the relays can cooperate with each other. Then, we provide three schemes to compress the observations at different relays and obtain several lower bounds to the bottleneck rate.
Numerical results show that with simple symbol-by-symbol relay processing and compression, the lower bounds obtained by the proposed achievable schemes can come close to the upper bound on a wide range of relevant system parameters.
Since the problem considered in this paper can be formulated either from \cite{4544988} (information transmission)
or from \cite{courtade2013multiterminal, estella2018distributed} (log-loss CEO), 
the proposed upper bound and achievable schemes apply to both models with different operational meanings.

\section{Problem Formulation}
\label{problem_formu}

\begin{figure}
	\centering
	\includegraphics[scale=0.50]{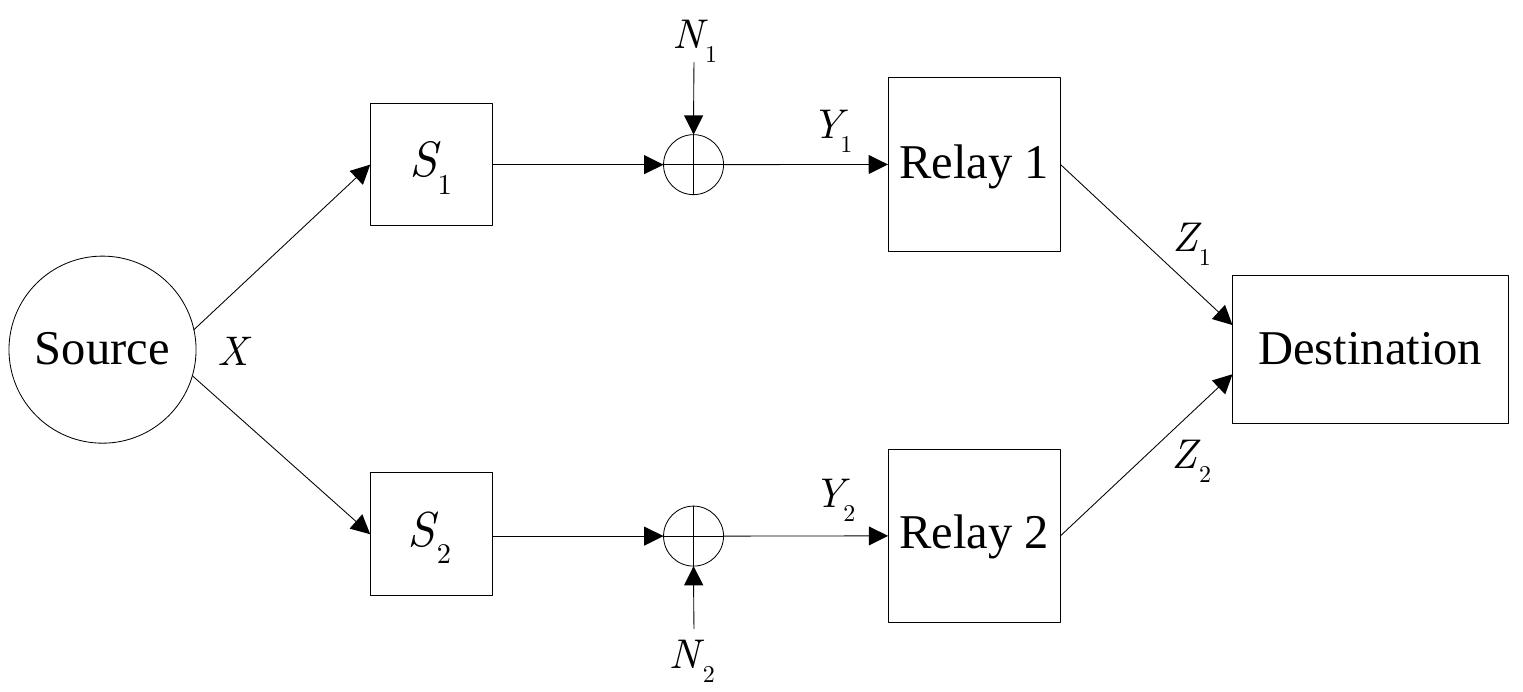}
	\vspace{-1em}
	\caption{A primitive Gaussian diamond channel with two relays.}
	\label{Block_diagram}
\end{figure}

As shown in Fig.~\ref{Block_diagram}, this paper considers a primitive Gaussian diamond channel with two relays and studies the distributed information bottleneck (D-IB) problem. 
The source node transmits signal $X$ to the relays over Gaussian channels with i.i.d. Rayleigh fading and each relay is connected to the destination via an error-free link with capacity $C_k, ~\forall~ k \in {\cal K} \triangleq \{ 1, 2 \}$. 
The observation of relay $k$ is
\begin{equation}\label{obser}
Y_k = S_k X + N_k,
\end{equation}
where $X \sim {\cal CN}(0, 1)$, $S_k \sim {\cal CN}(0, 1)$, and $N_k \sim {\cal CN}(0, \sigma^2)$ are respectively the channel input, channel fading from the source node to relay $k$, and Gaussian noise at relay $k$.

The relays are constrained to operate without knowledge of the codebooks, i.e., they perform oblivious processing and forward representations of their observations $Z_k$ to the destination.
According to \cite[Theorem~$1$]{4544988}, with the bottleneck constraints satisfied, the achievable communication rate at which the source node could encode its messages is upper bounded by the mutual information between $X$ and $Z_{\cal K} = \{Z_k\}_{k \in {\cal K}}$.
Hence, we consider the following D-IB problem
\begin{subequations}\label{IB_problem}
	\begin{align}
	\mathop {\max }\limits_{\{p(Z_k| Y_k, S_k)\}} & I(X; Z_{\cal K}) \label{IB_problem_a}\\
	\text{s.t.} \quad\;\;\; &  I(Y_{\cal T}, S_{\cal T}; Z_{\cal T}| Z_{{\cal T}^C}) \leq \sum_{k \in {\cal T}} C_k, ~\forall~ {\cal T} \subseteq {\cal K}, \label{IB_problem_b}
	\end{align}
\end{subequations}
where $C_k$ is the bottleneck constraint of relay $k$ and ${\cal T}^C$ is the complementary set of ${\cal T}$, i.e., ${\cal T}^C = {\cal K} \setminus {\cal T}$.
We call $I(X; Z_{\cal K})$ the bottleneck rate and $I(Y_{\cal T}, S_{\cal T}; Z_{\cal T}| Z_{{\cal T}^C})$ the compression rate.
Since the channel coefficient $S_k$ varies in each realization and is only known at the relay $k$, $S_{\cal T}$ is included in the compression rate formulation.
In (\ref{IB_problem}), we aim to find conditional distributions $p(Z_k| Y_k, S_k)$ such that collectively, the compressed signals at the destination preserve as much the original information from the source as possible.

Note that we formulate problem (\ref{IB_problem}) based on \cite[Theorem~$1$]{4544988}.
Alternatively, we may also arrive at (\ref{IB_problem}) by using \cite[Theorem~$1$]{courtade2013multiterminal} (with a simple change of variable).
For brevity, this paper describes the model using single letters.
The more operational model characterized with $n$-letter sequences can be similarly defined as in \cite{4544988} and \cite{courtade2013multiterminal}.
As explained in the introduction, \cite{4544988} and \cite{courtade2013multiterminal} respectively considered information transmission in uplink cloud radio access networks (CRANs) and distributed CEO problem, which have different operational meanings.
In a CEO problem, the source sequence is no longer considered as a codeword.
There is thus no need to assume obliviousness at the relays and $I(X; Z_{\cal K})$ is the relevant information (between the source sequence and the reconstructed sequence) rather than the communication rate \cite{estella2018distributed}.
In this sense, focusing on numerical problem (\ref{IB_problem}), the upper bound and achievable schemes proposed in the following sections apply to different scenarios.

\section{Informed Receiver Upper Bound}
\label{informed_ub}

Similar to the one-relay IB problems studied in \cite{caire2018information, info12040155, IBxu}, an obvious upper bound to problem (\ref{IB_problem}) can be obtained by assuming that the destination node knows all the channel coefficients $S_{\cal K} = \{S_k\}_{k \in {\cal K}}$.
We call this bound the informed receiver upper bound.
The D-IB problem then becomes
\begin{subequations}\label{IB_problem_ub}
	\begin{align}
	\mathop {\max }\limits_{\{p(Z_k| Y_k, S_k)\}} & I(X; Z_{\cal K}| S_{\cal K}) \label{IB_problem_ub_a}\\
	\text{s.t.} \quad\;\;\; &  I(Y_{\cal T}; Z_{\cal T}| Z_{{\cal T}^C}, S_{\cal K}) \leq \sum_{k \in {\cal T}} C_k, ~\forall~ {\cal T} \subseteq {\cal K}. \label{IB_problem_ub_b}
	\end{align}
\end{subequations}
Note that if $S_k, \forall k \in {\cal K}$ are fixed constants and are perfectly known at the destination node, condition $S_{\cal K}$ will be useless, and in this case, according to \cite[Theorem~$5$]{4544988}, the optimal value of problem (\ref{IB_problem_ub}) is 
\begin{align}\label{R_fixed_rho}
& R (\rho_{\cal K}, C_{\cal K}) = \nonumber\\
& \mathop {\max }\limits_{\{ r_k \}} \left\{\! \mathop {\min }\limits_{ {\cal T} \subseteq {\cal K}} \left\{\! \log\! \left[ 1 \!+\! \sum_{k \in {\cal T}^C} \rho_k \left( 1 \!-\! 2^{-r_k} \right) \right] \!+\! \sum_{k \in {\cal T}} (C_k \!-\! r_k) \!\right\} \!\right\}\!,
\end{align}
where $C_{\cal K} = \{C_k\}_{k \in {\cal K}}$, $\rho_{\cal K} = \{\rho_k\}_{k \in {\cal K}}$, $\rho_k = |S_k|^2/\sigma^2$ is the channel signal-to-noise ratio (SNR), and $r_k \geq 0$ is an intermediate variable.
Notice that if we formulate problem (\ref{IB_problem}) based on \cite[Theorem~$1$]{courtade2013multiterminal}, (\ref{R_fixed_rho}) can also be obtained by using \cite[Theorem~$3$]{courtade2013multiterminal} and \cite[Theorem~$2$]{estella2018distributed}.
Introducing an auxiliary variable $\beta$, $R (\rho_{\cal K}, C_{\cal K})$ can be  obtained by solving the following equivalent problem
\begin{subequations}\label{eq_problem}
	\begin{align}
	& \mathop {\max }\limits_{r_1, r_2, \beta} \; \beta \label{eq_problem_a}\\
	& \text{s.t.} \; \log\! \left[\! 1 \!+\! \sum_{k \in {\cal T}^C} \rho_k \left( 1 \!-\! 2^{-r_k} \right) \!\right] \!+\! \sum_{k \in {\cal T}} (C_k \!-\! r_k) \!\geq\! \beta, \forall~ {\cal T} \!\subseteq\! {\cal K}, \label{eq_problem_b}\\
	& \quad\quad 0 \leq r_k \leq C_k, ~\forall~ k \in {\cal K}. \label{eq_problem_c}
	\end{align}
\end{subequations}
It can be readily found that (\ref{eq_problem}) is a convex problem and can thus be optimally solved.

Using (\ref{R_fixed_rho}), problem (\ref{IB_problem_ub}), where the Rayleigh fading channels vary in each realization, can be solved by considering
\begin{subequations}\label{ergodic_problem}
	\begin{align}
	\mathop {\max }\limits_{c_{\cal K}} \quad & {\mathbb E} \left[ R (\rho_{\cal K}, c_{\cal K} ) \right] \label{ergodic_problem_a}\\
	\text{s.t.} \quad\; &  {\mathbb E} \left[ c_k (\rho_k) \right] \leq C_k, ~\forall~ k \in {\cal K}, \label{ergodic_problem_b}\\
	& c_k (\rho_k) \geq 0, ~\forall~ k \in {\cal K}, \label{ergodic_problem_c}
	\end{align}
\end{subequations}
where $c_{\cal K} = \{ c_k (\rho_k) \}_{k \in {\cal K}}$, $c_k (\rho_k)$ represents the allocation of the bottleneck rate $C_k$ for the channel realization with SNR $\rho_k$, and the expectation is taken over the random SNRs $\rho_{\cal K}$.
Note that though the optimal value of $R (\rho_{\cal K}, C_{\cal K})$ in (\ref{R_fixed_rho}) can be obtained by solving its equivalent and convex transformation (\ref{eq_problem}), its closed-form expression is unachievable.
Hence, different from problems \cite[(6)]{caire2018information} and \cite[(4)]{IBxu}, which admit the optimal closed-form solutions, problem (\ref{ergodic_problem}) is intractable.
We thus leave (\ref{ergodic_problem}) as an open problem for the future.

To obtain a simple upper bound to the bottleneck rate, besides the assumption that the destination node knows $S_{\cal K}$, we further assume that the relays can cooperate such that each relay also knows the observations $Y_k$ and $S_k$ of the other relay.
Actually, the network in this case can be seen as a system with a source node, a two-antenna relay, a destination node, and bottleneck constraint $C_1 + C_2$, and problem (\ref{IB_problem_ub}) becomes
\begin{subequations}\label{IB_problem_ub2}
	\begin{align}
	\mathop {\max }\limits_{p(Z_{\cal K}| Y_{\cal K}, S_{\cal K})} \quad & I(X; Z_{\cal K}| S_{\cal K}) \label{IB_problem_ub2_a}\\
	\text{s.t.} \quad\quad\;\;\; &  I(Y_{\cal K}; Z_{\cal K}| S_{\cal K}) \leq C_1 + C_2. \label{IB_problem_ub2_b}
	\end{align}
\end{subequations}
Denote vector $\bm S = (S_1, S_2)^T$. 
Obviously, the matrix $\bm S \bm S^H$ has only one positive eigenvalue $\lambda$.
It is known from \cite[(A17)]{IBxu} that the probability density function (pdf) of $\lambda$ is
\begin{equation}\label{pdf_lambda}
f_{\lambda} (\lambda) = \lambda e^{-\lambda}, ~\forall~ \lambda \geq 0.
\end{equation}
Then, according to \cite[Theorem~1]{IBxu}, the solution of problem (\ref{IB_problem_ub2}), which forms an upper bound to the bottleneck rate $I(X; Z_{\cal K})$ in (\ref{IB_problem_a}), is given by
\begin{equation}\label{R_up_KM}
R^{\text {ub}} = \int_{\nu \sigma^2}^{\infty} \left[ \log \left(1 + \frac{\lambda}{\sigma^2} \right) - \log (1 + \nu)\right] f_\lambda (\lambda) d \lambda,
\end{equation}
where $\nu$ is chosen such that the following bottleneck constraint is met
\begin{equation}\label{bottle_constr_KM}
\int_{\nu \sigma^2}^{\infty} \left( \log \frac{\lambda}{\nu \sigma^2} \right) f_\lambda (\lambda) d \lambda = C_1 + C_2.
\end{equation}

\section{Achievable Schemes}
\label{achiev_schems}

In this section, we provide several achievable schemes where each scheme satisfies the bottleneck constraint and gives a lower bound to the bottleneck rate.

\subsection{Quantized channel inversion (QCI) scheme}
\label{QCI_scheme}

In our first scheme, each relay first gets an estimate of the channel input using channel inversion and then transmits the quantized noise levels as well as the compressed noisy signal to the destination node.

In particular, using channel inversion to $Y_k$, i.e., multiplying $Y_k$ by $\frac{S_k^*}{|S_k|^2}$, we get
\begin{equation}\label{X_tilde}
{\tilde X}_k = X + \frac{S_k^*}{|S_k|^2} N_k \triangleq X + \sqrt{\xi_k} N_k',
\end{equation}
where $\xi_k = |S_k|^{-2}$, $N_k' = e^{-j \phi_k} N_k$, and $\phi_k$ denotes the phase of channel state $S_k$.
Due to the fact that the noise $N_k$ is rotationally invariant, $N_k'$ has the same statistics as $N_k$, i.e., $N_k' \sim {\cal CN}(0, \sigma^2)$.

We fix a finite grid of $J$ positive quantization points ${\cal B} = \{ b_1, \cdots, b_J \}$, where $b_1 \leq b_2 \leq \cdots \leq b_{J-1} < b_J$, $b_J = + \infty$, and define the following ceiling operation
\begin{equation}\label{ceiling}
\ceil[\big]{\xi_k}_{\cal B} = \min_{b \in {\cal B}} \{ \xi_k \leq b  \}.
\end{equation}
Then, each relay forces the channel (\ref{X_tilde}) to belong to a finite set of quantized levels by adding artificial noise, i.e., by introducing physical degradation as follows
\begin{align}\label{X_hat}
{\hat X}_k & = {\tilde X}_k + \sqrt{\ceil[\big]{\xi_k}_{\cal B} - \xi_k} N_k'' \nonumber\\
& = X + \sqrt{\xi_k} N_k' + \sqrt{\ceil[\big]{\xi_k}_{\cal B} - \xi_k} N_k'',
\end{align}
where $N_k'' \sim {\cal CN}(0, \sigma^2)$ is independent of everything else.
Since the relay $k$ knows $\xi_k$ in each channel realization, (\ref{X_hat}) is a Gaussian channel with noise power $\ceil[\big]{\xi_k}_{\cal B} \sigma^2$.
To evaluate the bottleneck rate, we denote the quantized SNR of channel (\ref{X_hat}) when $\ceil[\big]{\xi_k}_{\cal B} = b_{j_k}$ by
\begin{equation}\label{rho_hat}
{\hat \rho}_{k, j_k} = \frac{1}{b_{j_k} \sigma^2}, ~\forall~ k \in {\cal K}, j_k \in {\cal J},
\end{equation}
where ${\cal J} = \{ 1, \cdots, J \}$, and define probability
\begin{equation}\label{P_jk}
{\hat P}_{k, j_k} = {\text {Pr}} \left\{ \ceil[\big]{\xi_k}_{\cal B} = b_{j_k} \right\}, ~\forall~ j_k \in {\cal J}.
\end{equation}
From \cite[Theorem 5.3.1]{cover2012elements} it is known that the minimum number of quantization bits necessary for compressing $\ceil[\big]{\xi_k}_{\cal B}$ is
\begin{equation}\label{H_k}
{\hat H}_k = - \sum_{j_k = 1}^J {\hat P}_{k, j_k} \log {\hat P}_{k, j_k},
\end{equation}
which is actually the entropy of $\ceil[\big]{\xi_k}_{\cal B}$.
The remaining capacity available at relay $k$ for transmitting ${\hat X}_k$ is thus $C_k - {\hat H}_k$.
We use $c_{k, j_k}$ to denote the partial bottleneck rate allocated by relay $k$ to compress ${\hat X}_k$ for a given channel use with $\ceil[\big]{\xi_k}_{\cal B} = b_{j_k}$.
In addition, we use $R_{j_1, j_2}$ to indicate the achievable rate when $\ceil[\big]{\xi_1}_{\cal B} = b_{j_1}$ and $\ceil[\big]{\xi_2}_{\cal B} = b_{j_2}$.

Note that from (\ref{rho_hat}) and the definition of quantization points in $\cal B$, it is known that if $j_k = J$, ${\hat \rho}_{k, j_k} = 0$.
In this case, we let $c_{k, J} = 0$.
To evaluate the bottleneck rate, we first consider several special cases with $0$ SNR at relay~$1$ or relay~$2$ or both of them.
If $j_1 = j_2 = J$, it is obvious that $R_{J, J} = 0$.
If $j_1 = J$ and $j_2 \leq J-1$, the system reduces to an one-relay case as in \cite{caire2018information, IBxu, info12040155}. 
Then, using the bottleneck rate of the one-relay block-fading Gaussian channel given in \cite{winkelbauer2014rate}, we have the following achievable rate
\begin{equation}\label{R_j1j2_1}
R_{J, j_2} =  \log \left(1 + {\hat \rho}_{2, j_2}\right) - \log \left(1 + {\hat \rho}_{2, j_2} 2^{- c_{2, j_2} } \right).
\end{equation}
Similarly, if $j_1 \leq J-1$ and $j_2 = J$, we have $c_{2, J} = 0$ and
\begin{equation}\label{R_j1j2_2}
R_{j_1, J} =  \log \left(1 + {\hat \rho}_{1, j_1}\right) - \log \left(1 + {\hat \rho}_{1, j_1} 2^{- c_{1, j_1} } \right).
\end{equation}	
For the other cases with $j_1, j_2 \in {\cal J} \setminus J$, $R_{j_1, j_2}$ can be obtained as follows by using (\ref{R_fixed_rho}),
\begin{align}\label{R_lb_QCI_gene}
R_{j_1, j_2} & = \mathop {\max }\limits_{\{ r_{k, j_1, j_2} \}}\! \left\{ \mathop {\min }\limits_{ {\cal T} \subseteq {\cal K}} \!\left\{\! \log\! \left[\! 1 \!+\! \sum_{k \in {\cal T}^C} {\hat \rho}_{k, j_k} \left( 1 \!-\! 2^{-r_{k, j_1, j_2}} \right) \!\right] \right.\right. \nonumber\\
& \left.\left. + \sum_{k \in {\cal T}} \left( c_{k, j_k} - r_{k, j_1, j_2} \right) \right\} \right\}, ~\forall~ j_1,~ j_2 \in {\cal J} \setminus J.
\end{align}
Based on (\ref{R_j1j2_1}), (\ref{R_j1j2_2}), and (\ref{R_lb_QCI_gene}), a lower bound to the bottleneck rate, which we will denote by $R^{\text {lb}1}$, can be obtained by solving the following problem
\begin{subequations}\label{problem_QCI}
	\begin{align}
	\!\!\!\mathop {\max }\limits_{\left\{ c_{k, j_k} \right\}}  & \sum_{j_1=1}^J \sum_{j_2=1}^J {\hat P}_{1, j_1} {\hat P}_{2, j_2} R_{j_1, j_2} \label{problem_QCI_a}\\
	\text{s.t.} \quad &  \sum_{j_k=1}^{J-1} {\hat P}_{k, j_k} c_{k, j_k} \leq C_k - {\hat H}_k, ~\forall~ k \in {\cal K}, \label{problem_QCI_b}\\
	& c_{k, j_k} \geq 0, ~\forall~ k \in {\cal K},~ j_k \in {\cal J} \setminus J, \label{problem_QCI_c}\\
	& c_{k, J} = 0, ~\forall~ k \in {\cal K}. \label{problem_QCI_d}
	\end{align}
\end{subequations}
Due to the embedded $\max \min$ problem in (\ref{R_lb_QCI_gene}), it is difficult to directly solve (\ref{problem_QCI}).
However, similar to (\ref{eq_problem}), we may introduce $\beta_{j_1, j_2}$ for each $R_{j_1, j_2}, \forall j_1, j_2 \in {\cal J} \setminus J$ and rewrite (\ref{problem_QCI}) in a simple and convex form, which can then be optimally solved by some general tools.
Due to space limitation, we do not give the details here.

\subsection{Truncated channel inversion (TCI) scheme}
\label{TCI_scheme}

In the second scheme, we put a threshold $S_{\text {th}}$ on magnitude of the state $S_k$ such that zero capacity is allocated to states with $|S_k| < S_{\text {th}}$.
Specifically, when $|S_k| < S_{\text {th}}$, the relay $k$ does not transmit its observation, while when $|S_k| \geq S_{\text {th}}$, it takes ${\tilde X}_k$ in (\ref{X_tilde}) as the new observation and transmits a compressed version of ${\tilde X}_k$ to the destination node.
The information about whether to transmit the observation or not is encoded and sent to the destination node. 
Before evaluating the bottleneck rate, we first define the following probabilities
\begin{equation}\label{Prob_TCI}
{\tilde P}_k = {\text {Pr}} \left\{ |S_k| \geq S_{\text {th}} \right\}, ~\forall~ k \in {\cal K},
\end{equation}
and denote
\begin{align}\label{sigma_rho_tilde}
{\tilde H}_k & = - {\tilde P}_k \log {\tilde P}_k - (1 - {\tilde P}_k) \log (1 - {\tilde P}_k), \nonumber\\
{\tilde \sigma}_k^2 & = {\mathbb E} \left[ |S_k|^{-2} \sigma^2|~ |S_k| \geq S_{\text {th}} \right],~ {\tilde \rho}_k = \frac{1}{{\tilde \sigma}_k^2}, ~\forall~ k \in {\cal K},
\end{align}
where ${\tilde H}_k$ is the minimum number of bits required for informing the destination node if $|S_k| \geq S_{\text {th}}$ or not, ${\tilde \sigma}_k^2$ can be seen as the noise power in (\ref{X_tilde}) when $|S_k| \geq S_{\text {th}}$ and ${\tilde \rho}_k$ can be taken as the SNR.
When $|S_k| \geq S_{\text {th}}$, define an auxiliary variable
\begin{align}\label{X_tilde_g}
{\tilde X}_{k, g} = X + N_{k,g}',
\end{align}
where $N_{k, g}'$ is the Gaussian noise with zero-mean and the same second moment as $\sqrt{\xi_k} N_k'$ in (\ref{X_tilde}), i.e, $N_{k, g}' \sim {\cal CN}(0, {\tilde \sigma}_k^2)$.
Note that for a given threshold $S_{\text {th}}$, ${\tilde \sigma}_k^2$ is fixed.
It can thus be assumed to be known at the destination node with no bandwidth cost. 
Let $Z_{k, g}$ be a representation of ${\tilde X}_{k, g}$ given bottleneck constraint $C_k$.

Now we evaluate the bottleneck rate.
Since there are two relays and each of them determines to transmit or not based on the state magnitude, in the following, we consider four different cases by comparing $|S_k|$ with the threshold $S_{\text {th}}$ and derive a lower bound to the bottleneck rate for each case.
First, when $|S_1| < S_{\text {th}}$ and $|S_2| < S_{\text {th}}$, it is obvious that 
\begin{equation}\label{case1}
I (X; Z_1, Z_2| |S_1| < S_{\text {th}}, |S_2| < S_{\text {th}}) = 0,
\end{equation}
since both the relays do not transmit any observation to the destination node.
If $|S_1| \geq S_{\text {th}}$ and $|S_2| < S_{\text {th}}$, the system reduces to a one-relay case.
Using the bottleneck rate of the one-relay Gaussian channel in \cite{caire2018information} and the fact that for a Gaussian input, Gaussian noise minimizes the mutual information \cite[(9.178)]{cover2012elements}, $I (X; Z_1, Z_2| |S_1| \geq S_{\text {th}}, |S_2| < S_{\text {th}})$ can be lower bounded by
\begin{align}\label{R_lb_TCI10}
R_{1,0}^{\text {lb}2} & = I (X; Z_{1, g}, Z_{2, g}| |S_1| \geq S_{\text {th}}, |S_2| < S_{\text {th}}) \nonumber\\
& = \log \left(1 + {\tilde \rho}_1\right) - \log \Big(1 + {\tilde \rho}_1 2^{-\frac{C_1 - {\tilde H}_1}{{\tilde P}_1}} \Big),
\end{align}
where $Z_{k, g}$ is a representation of ${\tilde X}_{k, g}$ given in (\ref{X_tilde_g}).
Analogously, when $|S_1| < S_{\text {th}}$ and $|S_2| \geq S_{\text {th}}$, $I (X; Z_1, Z_2| |S_1| < S_{\text {th}}, |S_2| \geq S_{\text {th}})$ is lower bounded by
\begin{equation}\label{R_lb_TCI01}
R_{0,1}^{\text {lb}2} =  \log \left(1 + {\tilde \rho}_2\right) - \log \Big(1 + {\tilde \rho}_2 2^{-\frac{C_2 - {\tilde H}_2}{{\tilde P}_2}} \Big).
\end{equation}
When $|S_1| \geq S_{\text {th}}$ and $|S_2| \geq S_{\text {th}}$, a lower bound to $I (X; Z_1, Z_2| |S_1| \geq S_{\text {th}}, |S_2| \geq S_{\text {th}})$, can be obtained from (\ref{R_fixed_rho}) by replacing $\rho_k$ and $C_k$ with ${\tilde \rho}_k$ and $\frac{C_k - {\tilde H}_k}{{\tilde P}_k}$, i.e.,
\begin{align}\label{R_lb_TCI11}
& R_{1,1}^{\text {lb}2} = \nonumber\\
& \mathop {\max }\limits_{\{ r_k \}} \! \left\{\!\! \mathop {\min }\limits_{ {\cal T} \subseteq {\cal K}} \!\left\{ \!\log\!\! \left[\! 1 \!\!+\!\! \sum_{k \in {\cal T}^C} \!\! {\tilde \rho}_k \left(\! 1 \!-\! 2^{-r_k} \!\right) \!\right]\! \!+\! \sum_{k \in {\cal T}}\!\! \left(\! \frac{C_k \!-\! {\tilde H}_k}{{\tilde P}_k} \!-\! r_k \!\!\right) \!\!\right\} \!\!\right\}\!.
\end{align}
Accordingly, a lower bound to $I (X; Z_1, Z_2)$ is thus given by
\begin{equation}\label{R_lb_TCI}
R^{\text {lb}2} = {\tilde P}_1 (1 - {\tilde P}_2) R_{1,0}^{\text {lb}2} + (1 - {\tilde P}_1) {\tilde P}_2 R_{0,1}^{\text {lb}2} + {\tilde P}_1 {\tilde P}_2 R_{1,1}^{\text {lb}2},
\end{equation}
the value of which could be obtained by introducing an auxiliary variable to (\ref{R_lb_TCI11}) and solving the resulted convex problem as we did in (\ref{eq_problem}).

\subsection{MMSE-based scheme}
\label{MMSE_scheme}

In this subsection, we assume that each relay $k$ first produces the MMSE estimate of $X$ based on $(Y_k, S_k)$, and then source-encodes this estimate.
In particular, given $(Y_k, S_k)$, the MMSE estimate of $X$ obtained by relay $k$ is
\begin{equation}\label{x_bar_k}
{\bar X}_k = \frac{S_k^*}{|S_k|^2 + \sigma^2} Y_k = \frac{|S_k|^2}{|S_k|^2 + \sigma^2} X + \frac{S_k^*}{|S_k|^2 + \sigma^2} N_k.
\end{equation}
Taking ${\bar X}_k$ as a new observation, we assume that relay $k$ quantizes ${\bar X}_k$ by choosing $P_{Z_k| {\bar X}_k}$ to be a conditional Gaussian distribution, i.e., 
\begin{equation}\label{z_bar_k}
Z_k = {\bar X}_k + Q_k \triangleq U_k X + W_k,
\end{equation}
where $Q_k \sim {\cal CN}(0, D_k)$ is independent of everything else, $U_k = \frac{|S_k|^2}{|S_k|^2 + \sigma^2}$, and $W_k = \frac{S_k^*}{|S_k|^2 + \sigma^2} N_k + Q_k$.
Let ${\bar X}_{k, g}$ denote a zero-mean circularly symmetric complex Gaussian random variable with the same second moment as ${\bar X}_k$, i.e., ${\bar X}_{k, g} \sim {\cal {CN}} \left(0, {\mathbb E} \left[ |{\bar X}_k|^2 \right] \right)$, and $Z_{k, g} = {\bar X}_{k, g} + Q_k$.
Then, using the fact that Gaussian input maximizes the mutual information of a Gaussian additive noise channel, we have
\begin{equation}\label{mutual_x_bar_z}
I({\bar X}_k; Z_k) \leq I({\bar X}_{k, g}; Z_{k, g}) = \log \left( 1 + \frac{{\mathbb E} \left[ |{\bar X}_k|^2 \right]}{D_k} \right).
\end{equation}
Let 
\begin{equation}
\log \left( 1 + \frac{{\mathbb E} \left[ |{\bar X}_k|^2 \right]}{D_k} \right) = C_k.
\end{equation}
We thus have
\begin{equation}\label{mutual_XkbarZk_Ck}
I({\bar X}_k; Z_k) \leq C_k,
\end{equation}
and $D_k$ can be calculated as
\begin{equation}\label{D_k}
D_k = \frac{{\mathbb E} \left[ |{\bar X}_k|^2 \right]}{2^{C_k} - 1}.
\end{equation}

In the following, we first show that with (\ref{mutual_XkbarZk_Ck}), the bottleneck constraint of the considered system, i.e., 
\begin{equation}\label{bt_constr}
I({\bar X}_{\cal T}; Z_{\cal T}| Z_{{\cal T}^C}) \leq \sum_{k \in {\cal T}} C_k, ~\forall~ {\cal T} \subseteq {\cal K},
\end{equation}
can be guaranteed, and then provide a lower bound to the bottleneck rate $I(X; Z_1, Z_2)$.

Since $Z_1$ is independent of $Z_2$ given ${\bar X}_1$ and conditioning reduces differential entropy,
\begin{align}\label{mutual_X1barZ1_Z2}
I({\bar X}_1; Z_1| Z_2) & = h(Z_1| Z_2) - h(Z_1| {\bar X}_1, Z_2)\nonumber\\
& \leq h(Z_1) - h(Z_1| {\bar X}_1) = I({\bar X}_1; Z_1) \leq C_1.
\end{align}
Analogously, we also have
\begin{equation}\label{mutual_X2barZ2_Z1}
I({\bar X}_2; Z_2| Z_1) \leq I({\bar X}_2; Z_2) \leq C_2.
\end{equation}
Using the chain rule of mutual information,
\begin{align}\label{mutual_X1barX2bar_Z1Z2}
& I({\bar X}_1, {\bar X}_2; Z_1, Z_2) \nonumber\\
= & I({\bar X}_1, {\bar X}_2; Z_1) + I({\bar X}_1, {\bar X}_2; Z_2| Z_1)\nonumber\\
= & I({\bar X}_1; Z_1) \!+\! I({\bar X}_2; Z_1| {\bar X}_1) \!+\! I({\bar X}_1; Z_2| Z_1) \nonumber\\
+ & I({\bar X}_2; Z_2| {\bar X}_1, Z_1) = I({\bar X}_1; Z_1) + I({\bar X}_2; Z_2| {\bar X}_1, Z_1),
\end{align}
where we used
\begin{align}\label{mutual_2}
I({\bar X}_2; Z_1| {\bar X}_1) & = I({\bar X}_2; Q_1) = 0, \nonumber\\
I({\bar X}_1; Z_2| Z_1) & = I(Q_1; Z_2| Z_1) = 0.
\end{align}
(\ref{mutual_2}) holds since $Q_1$ is independent of everything else.
Moreover,
\begin{align}\label{mutual_X2barZ2_X1barZ1}
I({\bar X}_2; Z_2| {\bar X}_1, Z_1) & = I({\bar X}_2; Z_2| {\bar X}_1)\nonumber\\
& = h(Z_2| {\bar X}_1) - h(Z_2| {\bar X}_1, {\bar X}_2)\nonumber\\
& = h(Z_2| {\bar X}_1) - h(Z_2| {\bar X}_2)\nonumber\\
& \leq h(Z_2) - h(Z_2| {\bar X}_2)\nonumber\\
& = I({\bar X}_2; Z_2).
\end{align}
Combining (\ref{mutual_XkbarZk_Ck}), (\ref{mutual_X1barX2bar_Z1Z2}), and (\ref{mutual_X2barZ2_X1barZ1}), we have
\begin{equation}\label{mutual_X1barX2bar_Z1Z2_2}
I({\bar X}_1, {\bar X}_2; Z_1, Z_2) \leq I({\bar X}_1; Z_1) + I({\bar X}_2; Z_2) \leq C_1 + C_2.
\end{equation}
From (\ref{mutual_X1barZ1_Z2}), (\ref{mutual_X2barZ2_Z1}), and (\ref{mutual_X1barX2bar_Z1Z2_2}), it is known that the bottleneck constraint (\ref{bt_constr}) is satisfied.

The next step is to evaluate $I(X; Z_1, Z_2)$.
\begin{align}\label{mutual_XZ1Z2}
& I(X; Z_1, Z_2)\nonumber\\
= & h(Z_1, Z_2) - h(Z_1, Z_2| X)\nonumber\\
\geq & h(Z_1, Z_2| S_1, S_2) - h(Z_1, Z_2| X)\nonumber\\
= & h(Z_1, Z_2| S_1, S_2) - h(Z_1| X) - h(Z_2| X),
\end{align}
where the last step holds $Z_2$ is independent of $Z_1$ given $X$.
Then, we evaluate the terms in (\ref{mutual_XZ1Z2}) separately.
Denote
\begin{equation}\label{var_W1}
{\text {Var}} (W_k| S_k) = \frac{U_k \sigma^2}{|S_k|^2 + \sigma^2} + D_k \triangleq V_k.
\end{equation}
Since $X$, $N_k$, and $Q_k$ are independent normal variables, given $(S_1, S_2)$, $Z_1$ and $Z_2$ are jointly Gaussian.
Hence,
\begin{align}\label{mutual_XZ1Z2_S1S2}
& h(Z_1, Z_2| S_1, S_2)\nonumber\\
= & {\mathbb E} \left[ \log (\pi e)^2 \det \left( \begin{bmatrix} U_1^2 & U_1 U_2\\ U_1 U_2 & U_2^2 \end{bmatrix}
+ \begin{bmatrix} V_1 & 0 \\ 0 & V_2 \end{bmatrix}
\right) \right].
\end{align}
Moreover, since Gaussian distribution maximizes the entropy over all distributions with the same variance \cite{el2011network}, we have
\begin{equation}\label{zk_X}
h(Z_k| X) \leq {\mathbb E} \left[ \log \pi e \left( {\text {Var}} (U_k) |X|^2 + {\mathbb E} \left[ V_k \right] \right) \right].
\end{equation}
Substituting (\ref{mutual_XZ1Z2_S1S2}) and (\ref{zk_X}) into (\ref{mutual_XZ1Z2}), a lower bound to $I(X; Z_1, Z_2)$ can be obtained as follows
\begin{align}\label{R_lb_MMSE}
R^{\text {lb}3} & = {\mathbb E} \left[ \log \det \left( \begin{bmatrix} U_1^2 & U_1 U_2\\ U_1 U_2 & U_2^2 \end{bmatrix}
+ \begin{bmatrix} V_1 & 0 \\ 0 & V_2 \end{bmatrix}
\right) \right] \nonumber\\
& - \sum_{k=1}^K {\mathbb E} \left[ \log \left( {\text {Var}} (U_k) |X|^2 + {\mathbb E} \left[ V_k \right] \right) \right].
\end{align}

\section{Numerical Results}
\label{simulation}

In this section, we investigate the lower bounds obtained by the proposed achievable schemes and compare them with the upper bound.
For convenience, we assume equal bottleneck constraint, i.e., $C_1 = C_2 = C$, and when performing the QCI scheme, we choose the quantization levels as quantiles such that we obtain the uniform pmf ${\hat P}_{k, j_k} = \frac{1}{J}, \forall k \in {\cal K}, j_k \in {\cal J}$.
When performing the TCI scheme, we vary $S_{\text {th}}$ from $0$ to $2$ in step $0.1$ and choose the one which gives the largest rate value.

\begin{figure}
	\centering
	\includegraphics[scale=0.50]{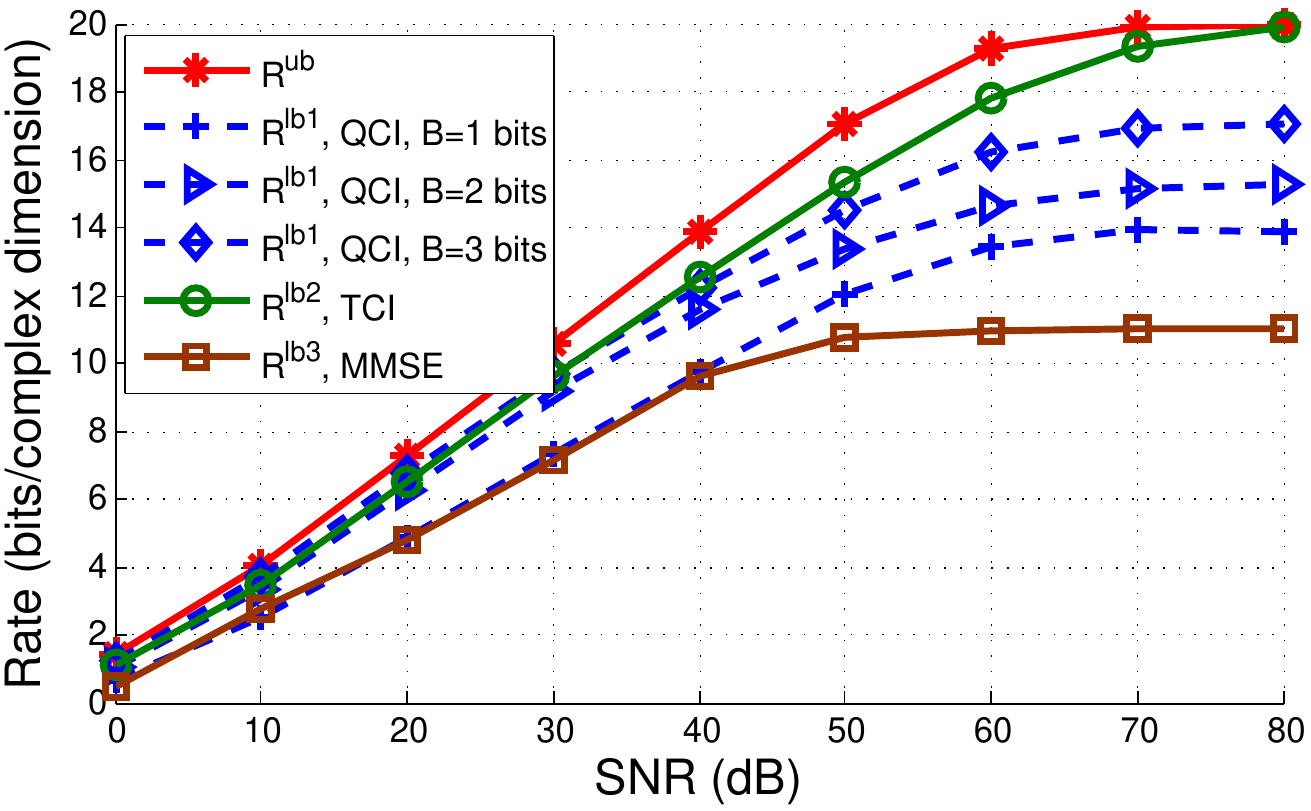}
	\vspace{-1em}
	\caption{Upper and lower bounds to the bottleneck rate versus $\rho$ with $C = 10$ bits/complex dimension.}
	\label{R_VS_rho}
\end{figure}

\begin{figure}
	\centering
	\includegraphics[scale=0.50]{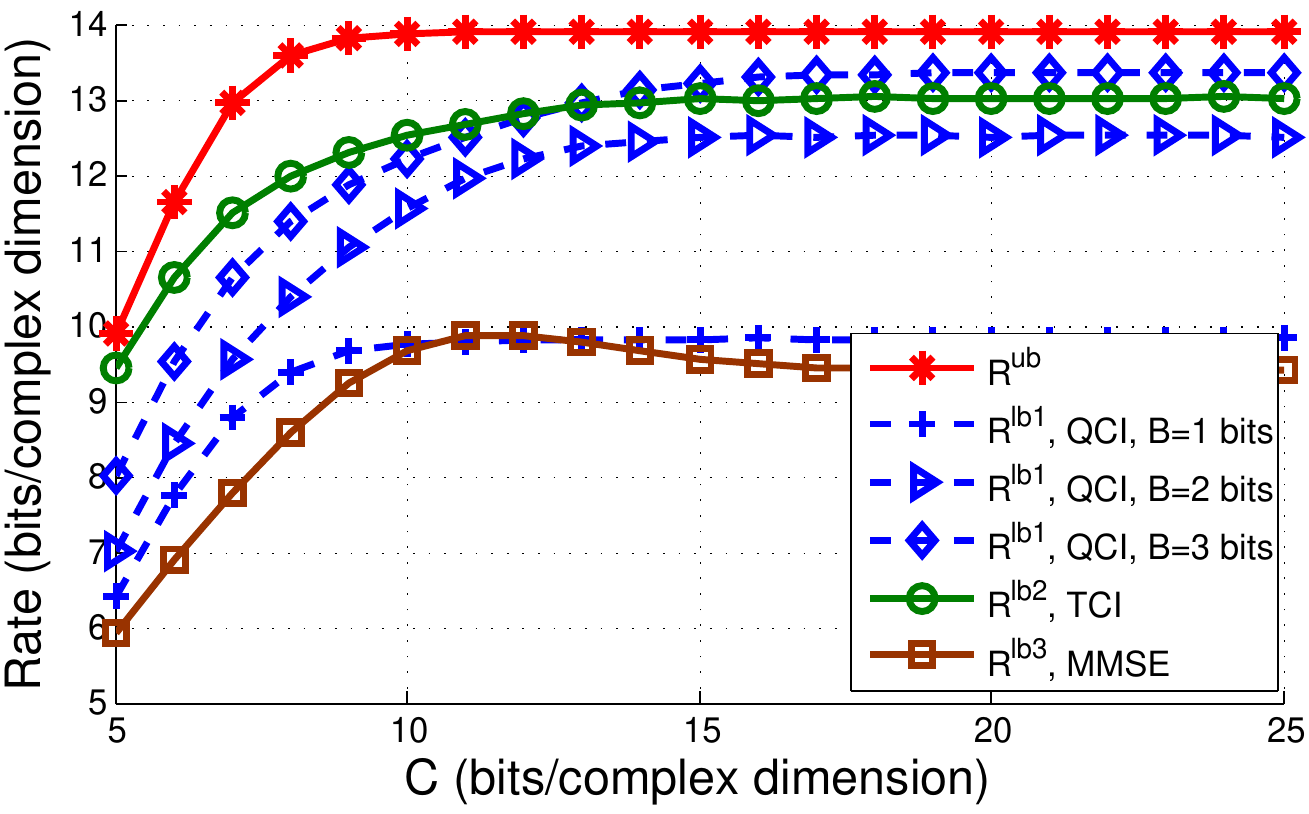}
	\vspace{-1em}
	\caption{Upper and lower bounds to the bottleneck rate versus $C$ with $\rho = 40$~dB.}
	\label{R_VS_C}
\end{figure}

In Fig.~\ref{R_VS_rho}, the upper and lower bounds are depicted versus SNR $\rho = \frac{1}{\sigma^2}$.
It can be found that for relatively small $\rho$, $R^{\text {lb}1}$ obtained by the QCI scheme with $3$ quantization bits and $R^{\text {lb}2}$ resulted from the TCI scheme get quite close to the upper bound.
As $\rho$ grows, both $R^{\text {ub}}$ and $R^{\text {lb}2}$ approach the sum capacity of the two relay-destination links, i.e., $C_1 + C_2$.
In addition, for the QCI scheme, there is a non-trivial optimal number of quantization bits which in general depends on the system parameters.

The effect of constraint $C$ is investigated in Fig.~\ref{R_VS_C}.
It shows that as $C$ increases, except $R^{\text {lb}3}$, all bounds grow monotonically and converge to constants.
When $C$ is large, $R^{\text {lb}1}$ essentially matches the $R^{\text {ub}}$, suggesting a good performance of the QCI scheme. 
Counterintuitively, $R^{\text {lb}3}$ increases first and then slightly decreases with $C$.
This is because when calculating $R^{\text {lb}3}$, we made two relaxations in (\ref{mutual_XZ1Z2}) and (\ref{zk_X}).

\section{Conclusions}
\label{conclusion}

This work extends the IB problem of the one-relay case in \cite{caire2018information} to a Gaussian diamond channel with Rayleigh fading.
Due to the bottleneck constraint, the destination node cannot get the perfect CSI from the relays.
Our results show that with simple symbol-by-symbol relay processing and compression, we can get bottleneck rate close to the upper bound on a wide range of relevant system parameters.
In the future, instead of assuming oblivious relays, we will address this primitive diamond channel with codebook knowledge at the relays.

\section*{Acknowledgments}
This work was supported by the European Union's Horizon 2020 Research and Innovation Programme under Marie
Skłodowska-Curie Grant No. 101024636 and the Alexander von Humboldt Foundation.
The work of S. Shamai has been supported by the European Union's Horizon 2020 Research and Innovation Programme with grant agreement No. 694630.

\bibliographystyle{IEEEtran}
\bibliography{IEEEabrv,Ref}
\end{document}